# Modeling the spillover risk of highly pathogenic avian influenza from wild birds to cattle in Denmark: A data-driven risk assessment framework

You Chang [a,*], Jose L. Gonzales [b], Erik Rattenborg [c], Mart C.M. de Jong [d], Beate Conrady [a]

[a] Department of Veterinary and Animal Sciences, University of Copenhagen, Frederiksberg C 1870, Denmark
[b] Department of Epidemiology, Bioinformatics and Animal Models, Wageningen Bioveterinary Research, Lelystad, the Netherlands
[c] SEGES Innovation P/S, Denmark
[d] Infectious Disease Epidemiology group, Department of Animal Sciences, Wageningen University and Research, the Netherlands

ABSTRACT

Since early 2024, highly pathogenic avian influenza virus (HPAIV) H5N1 of clade 2.3.4.4b has spilled over from wild birds to dairy cattle in the United States (U.S.), spreading to more than 1000 herds and threatening both animal and public health. Denmark's location along major migratory flyways and the lack of active HPAIV surveillance in cattle underscore the need to assess potential spillover risk from wild birds to cattle to strengthen preparedness. A quantitative spillover risk assessment model was developed to integrate data from Bird Flu Radar, eBird, and cattle density to estimate the weekly probability of HPAIV introduction from wild birds to cattle. The model was calibrated using observed U.S. spillover data and extrapolated to Denmark under the assumption of a comparable transmission rate parameter. Under the frequency-dependent model, the expected HPAIV introductions to Danish cattle via wild birds remain below 0.35 cases per week, with the highest temporal risk from December to March. High-risk areas were concentrated along the Danish coastline and near the German border. In contrast, applying a density-dependent model shifted the spatial risk toward regions with higher cattle densities, while the high-risk temporal periods remained unchanged. Overall, the results indicate a spatially and temporally variable risk of HPAIV spillover from wild birds to cattle in Denmark. The model establishes a data-driven framework to strengthen early warning and guide targeted surveillance efforts in high-risk regions.

## 1. Introduction

Since the first emergence of H5Nx goose/Guangdong lineage in wild birds in China (Xu et al., 1999), highly pathogenic avian influenza viruses (HPAIV) have spread globally, posing an increasing threat to both animal and human health. Early outbreaks in Asia were mainly limited to wild-to-domestic bird transmission, with only occasional spillover to humans (Claas et al., 1998). Clade 2.3.4.4 emerged in the early 2010s and drove large H5N8 and H5N2 epidemics in Europe and North America in 2014–2015 (Hill et al., 2017; Lee et al., 2015; Verhagen et al., 2015). This clade showed increased propensity for reassortment and lower virulence in some wild bird species, facilitating wider dissemination (Claes et al., 2016; Gass et al., 2023). More recently, the emergence of H5N1 clade 2.3.4.4b from 2020 onwards marked a new phase, with wider geographic spread, sustained global circulation in a broader host range across Europe, Africa, and Asia (Dupas et al., 2025; Sacristán et al., 2024).

The H5Nx HPAIV epidemic in Europe in 2020/2021 was among the most severe on record, resulting in 3777 detections and affecting approximately 22.9 million poultry across 31 countries (EFSA et al., 2022). By late 2021, the virus had spread to North America, first detected in Canada in December 2021 (Caliendo et al., 2022) and later in wild birds in North and South Carolina in January 2022 (Bevins et al., 2022), leading to the loss of over 100 million birds in the United States (U.S.) (USDA-APHIS, 2024). In addition to avian species, HPAIV spillovers have been increasingly reported in wild mammals, including seals, red foxes, and raccoons (Elsmo et al., 2023). Large-scale outbreaks of H5N1 in harbor seals in New England (USA) and in Quebec (Canada) marked the first mass mortality in mammals (Lair et al., 2024; Puryear et al., 2022).

In March 2024, H5N1 of the clade 2.3.4.4.b HPAIV was detected for the first time in U.S. dairy cattle, after being introduced by wild birds (Hu et al., 2024; Nguyen et al., 2025). Since then, the virus has spread to 1082 herds in 18 states, led to the culling of 166 million poultry, and been detected in 12,215 wild birds (USDA, 2025d). Infected cows typically show reduced appetite, fever, and decreased milk production (Sah et al., 2024; USDA-APHIS, 2024), with morbidity and mortality rates of approximately 20% and 2%, respectively (USDA, 2024a). About 70 human infections, primarily among poultry and dairy workers, have been reported (CDC, 2025). Transmission within and between dairy farms is most likely driven by cattle movements and indirect pathways such as contaminated milk and milking equipment, shared machinery, and farm personnel (Stone et al., 2024). In addition to cow-to-cow transmission, the virus has been detected in other animals near infected farms, including cats, raccoons, rodents, and nearby poultry flock (Mostafa et al., 2024). To mitigate the outbreak, U.S. authorities (USDA) have enforced negative-test requirements for interstate movement of lactating cows, enhanced biosecurity measures, and launched a national testing programme for unpasteurized milk (USDA, 2024b).

* Corresponding author.
*E-mail address:* you.chang@sund.ku.dk (Y. Chang).

Although HPAIV outbreaks in cattle have not yet been reported outside the U.S. (WOAH, 2025), several countries have launched surveillance and research efforts to assess spillover risk to cattle. Canada, for example, has tested pasteurized retail milk and raw bulk milk from dairy farms, with no evidence of HPAIV detected to date (CFIS, 2025). In Europe, retrospective serological investigations in cattle and other livestock species have been conducted in countries such as the Netherlands, Norway, Italy (Fabri et al., 2025; Fosse et al., 2025; IZSVe, 2024), generally reporting no evidence of past exposure. However, rare spillover events have been documented, including the first confirmed detection of clade 2.3.4.4b H5N1 in a ewe in Great Britain in 2025, likely associated with exposure to infected poultry on the same premises (Banyard et al., 2025). More recently, antibodies against HPAIV H5N1 were detected in a single dairy cow in the Netherlands, indicating past exposure without evidence of active infection or onward transmission (Wageningen Bioveterinary Research, 2026).

Denmark's location along major migratory flyways (Handberg et al., 2010) and its high-density livestock production with 12.6 dairy head/km2 (Landbrug and Fødevarer, 2024) making it particularly vulnerable to HPAIV introduction from wild birds. In recent years, Denmark has experienced recurrent HPAIV detections in wild birds and poultry, underscoring its vulnerability to cross-species transmission events (EFSA et al., 2021). Like many other European countries, Denmark has not implemented active cattle-specific surveillance, such as serological or milk-based testing, to monitor possible HPAIV exposure in cattle, relying instead on passive, suspicion-based reporting. This lack of active surveillance highlights the importance of quantitative risk assessments to identify where and when spillover to cattle might occur and to inform the design of targeted surveillance in high-risk areas.

Several tools have been developed to assess HPAIV spillover risk and support early warning systems, yet they primarily focus on wild birds and poultry. The U.S. spillover risk assessment framework (Prosser et al., 2024) and EFSA's Bird Flu Radar tool for EU (Gargallo et al., 2022), use wild bird abundance and prevalence data to predict the spillover risk and are validated against poultry outbreaks. However, they do not evaluate the risk of spillover to cattle, an increasingly pressing concern.

To address the lack of HPAIV surveillance in Danish cattle and the absence of cattle-specific spillover risk assessment tools, we developed a quantitative risk assessment model to estimate the weekly infection probability from wild birds to cattle. The model aims to identify where and when HPAIV spillover is most likely to occur in dairy cattle, thereby supporting early warning efforts and guiding targeted surveillance to inform farmers and authorities. This model does not quantify the consequences following successful spillover such as sustained farm-to-farm transmission among dairy farms.

## 2. Method

In this study, it was assumed that any dominantly circulating HPAIV genotype in wild birds is capable of spillover, given that U.S. cattle events involved distinct genotypes (USDA, 2025a, 2025b). This assumption is also supported by recent research showing that European viruses can replicate in the udder and transmit via contaminated milk, similar to U.S. strains (Bordes et al., 2024; Halwe et al., 2024) and evidence of spillover and presence of virus' RNA in milk from an infected ewe in the UK (Banyard et al., 2025). As the spillover to cattle events have only been reported in the U.S., we calibrated the transmission rate parameter from U.S. spillover events and extrapolated the parameter to the Danish context.

In detail, the Section 2.1 introduces the model framework for HPAIV introduction from wild birds to cattle. The variables and datasets are described in Section 2.2, followed by parameter description in Section 2.3. In Section 2.4, we describe the calibration of the model based on U.S. spillover events, followed by the extrapolation of the model to Danish data in Section 2.5. Lastly, a sensitivity analysis was conducted in Section 2.6.

### 2.1. Model framework for HPAIV introduction from wild birds to cattle

We developed a quantitative model to estimate the weekly probability of HPAIV introduction from wild birds to cattle at the grid-cell level. The model uses a frequency-dependent transmission assumption, which is commonly applied to between-individual transmission, while the density-dependent assumption is more commonly applied to between-farm scale (Lambert et al., 2023).

In the context of cross-species transmission, frequency-dependent transmission retains the same mathematical structure as in single-host transmission ($\frac{\beta \cdot I}{N}$) with $I$ representing the number of infectious individuals in the source species and $N$ representing the population size of the recipient species (Chang et al., 2023).

In the present model, the number of cattle ($Nc$) in the denominator reflects the assumption that the frequency of wild bird visits to cattle farms is determined by bird behavior and farm attractiveness (e.g. availability of feed or water), rather than by cattle density within a grid cell. Each bird visit therefore allows contact with only a limited number of cattle, such that exposure from wild birds is distributed among cattle present within the grid. This assumption is analogous to formulations widely used in vector-borne disease models, where vector biting rates are independent of host population size (May, 1991; Adams and Kapan, 2009).

However, alternative contact structures may also be considered. In particular, a density-dependent formulation implies that contacts between birds and cattle increase with cattle population size. This assumption was therefore examined in a sensitivity analysis (Section 2.6).

Under the frequency-dependent assumption, the weekly probability of spillover events within each geographical cell is given by $P$and can be written as:

$$P = 1 - e^{-\frac{\beta \cdot Prev \cdot A \cdot dt}{Nc}} \quad (1)$$

where $\beta$ is the transmission rate parameter from wild birds to cattle. $Prev$ is the prevalence of infected wild birds, $A$ represents the weekly wild bird abundance of selected waterfowl species (details in Section 2.2) and $Nc$ represents the number of dairy cattle. Grid cells with $Nc = 0$ were excluded from the analysis, as spillover to cattle cannot occur in the absence of cattle. The variable $dt$ represents the time interval, measured in weeks.

### 2.2. Variables and data

Building upon the theoretical model framework introduced in Section 2.1, this section describes the variables used to quantify spillover risk (Table 1) and the data sources used to estimate them (Table 2). Data were extracted from U.S and Denmark: The model was calibrated using observed U.S. spillover data (Section 2.4) and extrapolated to Denmark using Danish data (Section 2.5).

**Table 1**
Variables in the model.

| Symbol | Definition | Unit | Role in the model |
|---|---|---|---|
| ***A*** | Wild birds abundance (using a standardized relative index) | Index | Predictor |
| $Prev_{US}$ | Weekly prevalence of HPAIV in wild birds in the U.S. during outbreak weeks | Unitless (0–1) | Calibrated jointly with $\beta$ |
| $Prev_{DK}$ | Weekly prevalence of HPAIV in wild birds in Denmark | Unitless (0–1) | Predictor: Assumed conditionally on outbreak status |
| $Ind_{WB}$ | Indicator of HPAIV presence in wild birds in Denmark | (0 or 1) | Predictor: Outbreak indicator variable (from Bird Flu Radar) |
| ***Nc*** | Number of cattle per cell | Animal | Predictor: Exposure population |
| ***P*** | The probability of spillover events from wild birds to cattle | Unitless (0–1) | Observed in calibration; predictor |

**Table 2**
Overview of data and tools used in this study, including their associated spatial and temporal resolution.

| Data/tool names | Corresponding variable name | Spatial resolution | Temporal resolution and span |
|---|---|---|---|
| eBird | *A* | 9 × 9 km | On a weekly basis (period: 2022/2023) |
| Gridded Livestock of the World database | *Nc* (for model calibration) | 0.0833 × 0.0833 decimal degrees | 2015 |
| Danish Central Husbandry Register | *Nc* (for prediction) | point | 2024 |
| World Animal Health Information System | *Spillover cases* | Point | Weekly |
| Bird Flu Radar | $Ind_{wb}$ | 50 × 50 km | On a weekly basis (period: 2023–2025) |

### 2.2.1. Variables

The variables include wild bird abundance (*A*), the number of cattle (*Nc*), which serve as inputs for both model calibration and extrapolation. The prevalence of HPAIV in the U.S. wild birds during outbreaks ($Prev_{US}$) was estimated jointly with the transmission rate parameter $\beta$ during the model calibration. The outbreak status indicator ($Ind_{WB}$) was used to indicate the HPAIV presence in wild birds in Denmark, linking the Danish prevalence ($Prev_{DK}$) to the $Prev_{US}$ through a scaling parameter *r* (see detail Section 2.3). The observed spillover cases in the U.S. were used only for model calibration.

### 2.2.2. Data

Five datasets were extracted to quantify the variables in Table 1, including: (i) eBird Status and Trends (wild bird relative abundance), (ii) Gridded Livestock of the World database (cattle density for the U.S.), (iii) Danish Central Husbandry Register (cattle density for Denmark), (iv) World Animal Health Information System (HPAIV outbreaks report), and (v) Bird Flu Radar (HPAIV in wild bird indicator). Four of these datasets are publicly available except that the Danish cattle density data were obtained under restricted access. Each dataset is described below, along with a summary of how spatial and temporal resolutions were harmonized across the datasets.

*2.2.2.1. eBird status and trends.* The eBird Status and Trends dataset provides global, weekly predictions of relative abundance for 2408 bird species (Fink et al., 2013, 2020, 2023). Relative abundance is expressed as a standardized observation, after accounting for observer effort and detectability. While it is not an absolute population count, this index is proportional to real local abundance and captures spatial and temporal variation in bird presence in a way that is comparable across regions and time. In this study, the eBird relative abundance index was used as a quantitative measure of wild bird abundance (*A*) in both the calibration and extrapolation analyses.

The eBird datasets represent a typical annual cycle of wild bird abundance, that is generally stable across years. Predictions are generated using the AdaSTEM machine learning framework, which partitions space into variable-sized blocks that are modeled with boosted regression trees. Models incorporate 84 predictors, including observation efforts, seasonal trends, and environmental variables from remote sensing to capture bird–habitat associations across the diverse landscapes.

These data were used to quantify wild bird abundance variable (*A*) in both calibration (U.S.) and extrapolation (Denmark). The temporal and spatial distributions of wild bird abundance are presented in Supplementary Figures S1 and S2. The present study covers 54 Anseriformes species due to their key role in the spillover of HPAIV to livestock (see selection criteria in Supplementary Table S1).

*2.2.2.2. Dairy cattle density.* Cattle density from the U.S. was obtained from the Gridded Livestock of the World database (GLW v4), which collects and harmonizes global data on livestock distribution, including cattle. In this database, cattle numbers are disaggregated within each 0.0833° census polygon, using dasymetric method, which applies weights derived from Random Forest models based on high-resolution environmental predictor variables (Gilbert et al., 2018, 2022). As the GLW database provides total cattle density, we estimated dairy cattle density for each cell by applying a state-specific dairy-to-cattle ratio from the cattle inventory data (Munster et al., 2007; USDA, 2025c).

Dairy cattle density in Denmark was derived from the not public domain Danish Central Husbandry Registration (CHR), received by SEGES. Farm locations, herd size, and herd type in 2024 were extracted from CHR to estimate the number of dairy cattle per cell (see Supplementary Figure S3).

*2.2.2.3. World animal health information system.* Observed spillover events were extracted from the World Animal Health Information System (WAHIS) and supplemented with news reports (USDA, 2025d, 2025a, 2025b; WOAH, 2025). Only confirmed wild bird-to-cattle introductions were considered, while WAHIS outbreaks representing secondary farm-to-farm transmission were excluded. Three confirmed spillover events were identified: one in Texas (genotype B3.13) and two independent events in Nevada and Arizona (both genotype D1.1). The Texas event was reported in March 2024 (Nguyen et al., 2025), but by that time local spread had already occurred, and the exact index farm and introduction site cannot be determined. Therefore, only the Nevada and Arizona events were used for model calibration.

To construct the empirical weekly probability of spillover (*P*) for each grid cell, we counted the number of confirmed spillover events occurring within a two-month window (one month before and one month after each outbreak) and divided this by the number of cattle farms (*Nc*) in that cell. The observed probabilities were then used as the response variable in the calibration process described in Section 2.4.

*2.2.2.4. Bird Flu Radar.* The Bird Flu Radar tool estimates the weekly likelihood of HPAIV presence in European wild bird populations (Gargallo et al., 2022). It uses a scenario tree model that combines a bird abundance model, based on the AdaSTEM approach applied to EuroBirdPortal data, and a migration model using the Integrated Nested Laplace Approximation (INLA) method applied to the EURING DataBank data. Twelve waterfowl species are included (listed in Supplementary Table S1), representing 89% of available European ring recovery data. The output of the model is the weekly likelihood of HPAIV presence in European wild bird populations $Ind_{WB}$.

*2.2.2.5. Spatial and temporal resolution.* Extracted datasets with varying spatial and temporal resolutions (Table 2) were harmonized to a common resolution for analysis. The eBird spatial grids (9 × 9 km) were used as the spatial reference. To match this resolution, global cattle density data, originally at 0.0833 × 0.0833 decimal degrees, were resampled using bilinear interpolation. Danish cattle data, provided as point-level farm locations, were aggregated to a 9 × 9 km cell by summing the number of animals within each cell. For each grid cell, if its centroid falls within a Bird Flu Radar cell (50 × 50 km), the corresponding HPAIV presence indicator from Bird Flu Radar was assigned.

The temporal unit of analysis was at weekly level, consistent with the weekly resolution of both the eBird and the Bird Flu Radar datasets. The Bird Flu Radar data, available from May 2023 to August 2025, incorporates weekly real-time outbreak information. In contrast, global distribution on relative bird abundance and cattle density are updated less frequently; therefore, their distribution patterns were assumed stable over recent years (Table 2).

### 2.3. Parameter descriptions

The model includes the following epidemiological parameters:

- $\beta_{DK}{\cdot}Prev_{US}$: This parameter combines the transmission rate parameter from wild birds to cattle with the prevalence of HPAIV in the U. S. wild birds ($Prev_{US}$) during outbreak weeks. It was estimated during the model calibration (Section 2.4) using U.S. data on wild birds' abundance, dairy cattle density, and the weekly infection probability per geographical cell.
- $r$: The scaling parameter is defined as the ratio of the Danish to U.S. spillover rate coefficients: ($r = \frac{\beta_{DK}{\cdot}Prev_{DK}}{\beta_{US}{\cdot}Prev_{US}}$). Under the default assumption, $r$ is specified as:

$$r = \begin{cases} 1, & If\ Ind_{wb} = 1 \quad (outbreak\ period) \\ 0.05, & If\ Ind_{wb} = 0 \quad (non-outbreak\ period) \end{cases} \tag{2}$$

  During outbreak weeks in Denmark ($Ind_{wb} = 1$), we assume that the product of prevalence in Danish wild birds and the transmission rate parameter ($\beta_{DK}{\cdot}Prev_{DK}$) is comparable to that estimated under the U.S. outbreak conditions. In contrast, during non-outbreak weeks, we assume $r = 0.05$, reflecting evidence that wild bird prevalence is much lower outside epidemic peaks. Published studies show temporal variation, with outbreak periods often having an order of magnitude higher prevalence than non-outbreak periods (Gargallo et al., 2022; Munster et al., 2007; USDA and APHIS, 2025). These assumptions define the baseline scenario, and alternative values of $r$ were explored in the sensitivity analysis (see Section 2.6).
- *Threshold probability*: This parameter determines whether HPAIV is considered present in wild birds according to the Bird Flu Radar tool. As the Bird Flu Radar calculates the probability of HPAIV in wild birds, which is used to determine the HPAIV presence indicator $Ind_{wb}$ via a threshold probability. The threshold probability is optimized and validated against reported outbreaks from 2016 to 2021, identified an optimal threshold probability at 0.045: if the predicted probability exceeds this value, the wild bird indicator is $Ind_{wb} = 1$; otherwise, $Ind_{wb} = 0$. Alternative thresholds and the associated impact were explored in sensitivity analyses (see Section 2.6).

### 2.4. Model calibration based on U.S. spillover events

To calibrate the model in Eq.(1), we fitted a generalized linear model (GLM) with a binomial error distribution and a 'cloglog' link to U.S. spillover data, from which the transmission rate was estimated (Bouma et al., 1995; Gonzales et al., 2012; Velthuis et al., 2003). Parameter uncertainty was quantified using 95% confidence intervals based on the Wald method.

$$\log(-\log(1-P)) = \log\left(\frac{\beta_{US}{\cdot}Prev_{US}A{\cdot}dt}{Nc}\right) = \log(\beta_{US}{\cdot}Prev_{US}) + \log\left(\frac{A{\cdot}dt}{Nc}\right) \tag{3}$$

where $P$ is the observed probability of spillover events in a week derived in Section 2.2. The term $\log\left(\frac{A{\cdot}dt}{Nc}\right)$ is fitted as an offset variable with a fixed coefficient of 1. In standard formulations, prevalence data would also enter the offset, allowing direct estimation of $\beta_{US}$. Available HPAIV prevalence estimates are not suitable for our analysis: USDA data lack sufficient temporal and spatial resolution (USDA and APHIS, 2025), and a high-resolution model calibrated before the emergence of H5N1 clade 2.3.4.4 (Kent et al., 2022; Prosser et al., 2024), likely underestimates prevalence driving recent spillover to cattle. Consequently, we estimate the combined value $\beta_{US}{\cdot}Prev_{US}$, where $Prev_{US}$ represents the prevalence in wild birds during outbreaks in the U.S., when spillover to cattle events occurred.

### 2.5. Extrapolation of the model to Danish data

To extrapolate the model to Denmark, the value of $\beta_{US}{\cdot}Prev_{US}$ estimated in Section 2.4 was used and its associated uncertainty was propagated using the 95% confidence interval. To account for differences in HPAIV prevalence in wild birds between countries and between outbreak and non-outbreak periods, as well as potential variation in the transmission rate, a scaling parameter, $r$, was introduced into the weekly spillover probability equation:

$$P = 1 - e^{-\frac{(\beta_{US}\cdot Prev_{US})\cdot r\cdot A\cdot dt}{Nc}} \tag{4}$$

Here, $P$ represents the probability of spillover to an individual animal within a grid and week. The expected number of spillover cases in each grid cell and week is therefore obtained by summing individual level probabilities across cattle:

$$E[Spillover\ cases\ per\ week] = Nc{\cdot}P \tag{5}$$

Weekly expected spillover cases are summed across grid cells and over time to obtain the expected annual number spillover cases at the national level.

### 2.6. Sensitivity analyses

To assess the robustness of model predictions, we performed a parameter uncertainty analysis and a structural sensitivity analysis.

#### 2.6.1. Parameter uncertainty analysis

We assessed parameter uncertainty using a multi-parameter sensitivity analysis. Specifically, $\beta_{US}{\cdot}Prev_{US}$ was varied within its estimated confidence interval, while the *Threshold probability* and $r$ were simultaneously varied across a range of plausible values (Table 3). In this context, four values for the model parameter *Threshold probability* (0.015, 0.030, 0.045, 0.060) were selected, representing the optimized cut-offs for HPAIV presence indicator values in Bird Flu Radar tool. The corresponding sensitivity and specificity values of the Bird Flu Radar for each tested threshold probability are shown in Supplementary Table S2.

**Table 3**
Parameter ranges used in the sensitivity analysis.

| Parameter | Value |
|---|---|
| ***r*** | 0.50, 0.75, 1, 1.25, 1.50 |
| ***Threshold probability*** | 0.015, 0.030, 0.045, 0.060 |

The scaling parameter *r* during the outbreak was varied symmetrically by ± 25% and ± 50% around 1.0 to capture scenarios with lower or higher relative prevalence, reflecting differences in HPAIV prevalence among Danish and U.S. wild birds during outbreaks, as well as potential variation in transmission rate parameters between countries. Although the explored range of *r* was modest, we conducted a regression analysis to examine its relationship with the model output.

Combining the four *Threshold probability* values and five *r* values resulted in 20 parameter combinations, which were used to evaluate the influence of parameter uncertainty on outputs, including (i) the total number of predicted cases, (ii) the timing of the peak-risk week, and (iii) the spatial distribution of high-risk regions.

#### 2.6.2. Structural sensitivity analysis

To account for uncertainty in the model assumption, we also assessed an alternative transmission formulation. Specifically, the spillover probability equation based on the frequency-dependent formulation was replaced with a density-dependent formulation:

$$P = 1 - e^{-(\beta_{US_den}\cdot Prev_{US})\cdot r\cdot A\cdot dt} \quad (5)$$

The composite parameter $\beta_{US_den}\cdot Prev_{US}$ was calibrated using the U.S. spillover data, and predictions for Denmark were then recomputed under this model assumption.

## 3. Results

### 3.1. Spatial and temporal distribution of expected spillover cases

The weekly transmission rate from wild birds to cattle ($\beta_{DK}\cdot Prev_{DK}$) was estimated as $9.16 \times 10^{-6}$ per relative abundance per week (95% CI: $1.53 \times 10^{-6}$, $2.84 \times 10^{-5}$). Consequently, the expected number of spillover infections per cell per week in Denmark, ranges from 0 to $8.1 \times 10^{-3}$ cases, with a median of $2.8 \times 10^{-5}$ cases.

Temporal variation in the expected number of cattle infected by wild birds, aggregated across all grid cells in Denmark, is shown in Fig. 1. In 2024, the expected weekly spillover events were highest with 0.35 cases during calendar weeks 2–10 (95% CI: 0.06–1.1). For descriptive purpose, we define weeks with an expected number of spillover events exceeding 0.1 as periods of relative elevated spillover risk. Overall, 18 weeks had an expected case above 0.1 (weeks 1–15 and 49–51). During the first eight months of 2025, the highest expected number of spillover events occurred in weeks 10–12, with an estimated 0.25 cases (95% CI: 0.04–0.83). In total, nine weeks within this period had expected cases exceeding 0.1. Across Denmark, the cumulative expected number of spillover cases was estimated at 5.44 cases (95% CI: 0.90, 16.9) in 2024, and 1.92 (95% CI: 0.32, 5.94) for the first eight months of 2025.

The spatial distribution of spillover risk is heterogeneous, as shown by the predicted number of cattle infections by wild birds per cell during the first week of each month in 2024 (Fig. 2). In Denmark, the greatest risk of HPAIV spillover from wild birds to cattle is observed along coastlines, near inland water bodies, and in regions bordering Germany.

### 3.2. Parameter sensitivity analysis

As shown in the spillover probability equation (Eq.5), increasing the scaling parameter *r* led to a monotonic increase in the predicted spillover risk. When *r* was halved, the expected number of cattle cases decreased 0.48-fold from 5.44 to 2.62 cases, whereas increasing *r* to 1.5 raised the estimate 1.44-fold to 7.86 cases. The corresponding 95% confidence intervals ranged from 0.45 to 8.43 for $r = 0.5$ and 1.36–25.29 for $r = 1.5$ (Fig. 3A). Over the range of scaling parameter values considered, the expected number of spillover cases increased approximately linearly with the scaling parameter (adjusted $R^2 = 1$). This behavior arises because spillover probabilities remain small, such that the exponential formulation of Eq. (5) is well approximated by a linear relationship. As a result, expected outcomes can be approximately rescaled if future evidence suggests higher or lower scaling parameter. Despite variations in the absolute number of expected cases, the timing of high-risk weeks, defined as weeks within the top 25% of predicted spillover risk, remained stable, primarily falling between weeks 1 and 13.

Increasing the *Threshold probability* for HPAIV presence in wild birds led to a monotonic decrease in the predicted spillover risk. For example, when the *Threshold probability* was 0.015, the expected number of spillover cases was 1.2 times higher than at a threshold of 0.06 (Fig. 3). This pattern occurs because a higher threshold reduces the sensitivity of Bird Flu Radar, resulting in fewer weeks and spatial cells in Denmark being classified as HPAIV-positive in wild birds, and consequently lower predicted spillover cases. As shown in Fig. 3B, lowering the *Threshold probability* to 0.015 resulted in a modest shortening of the early-season high-risk period (from weeks 1–13–1–11) and a shift of high-risk weeks toward the end of the year (weeks 49–51).

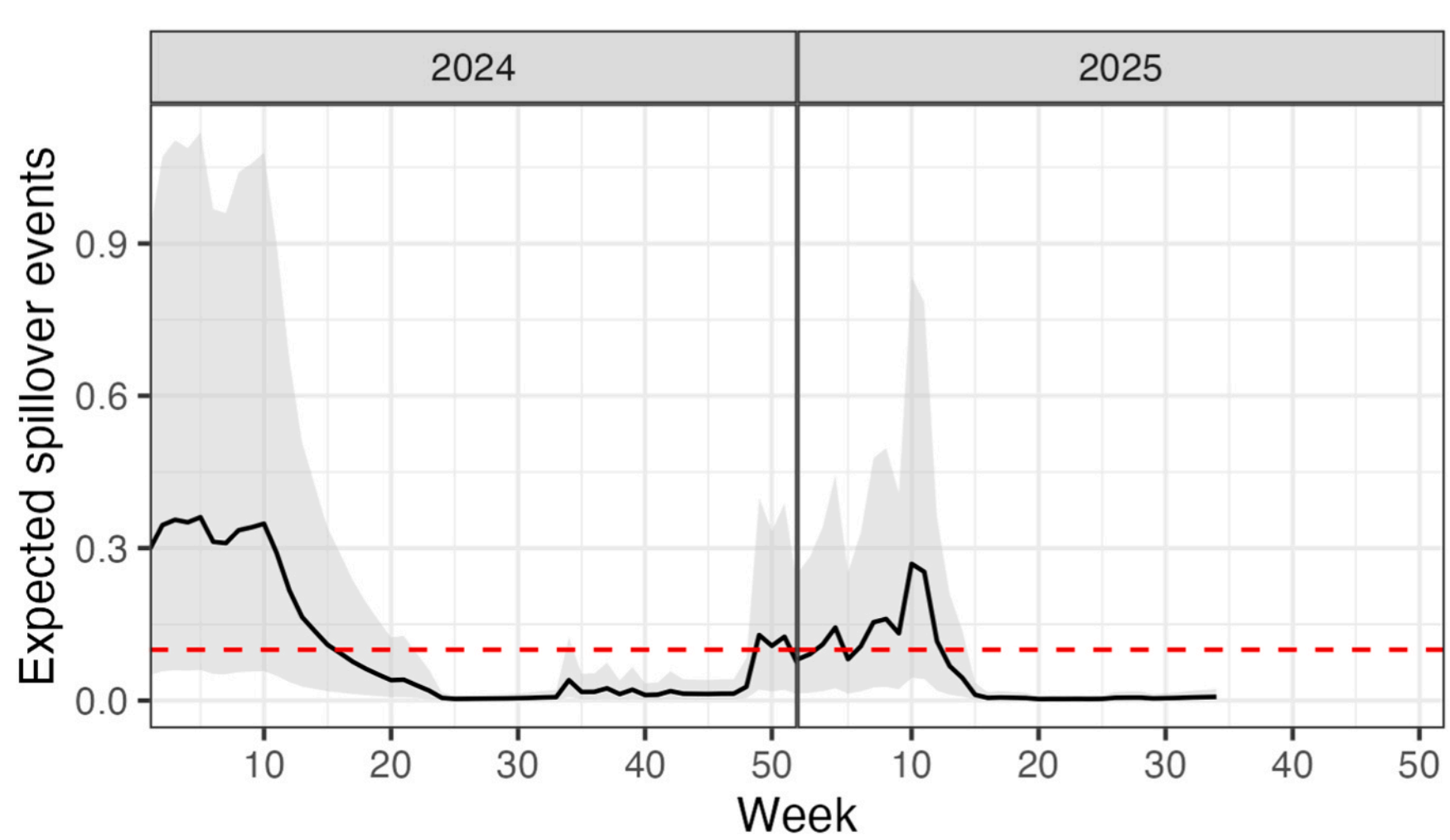

**Fig. 1.** Temporal changes of the expected spillover cases of HPAIV from wild birds to cattle ($Nc\cdot P\left(\frac{Spillover}{Nc}\right)$) in Denmark. The black line represents the predicted weekly number of expected spillover cases in Denmark under the baseline scenario. The grey shaded area represents the 95% confidence interval of predictions. The red dotted line indicates a threshold of 0.1 expected cases per week, used to highlight periods of relatively elevated spillover risk. Data for 2025 were only up to August at the time of data extraction.

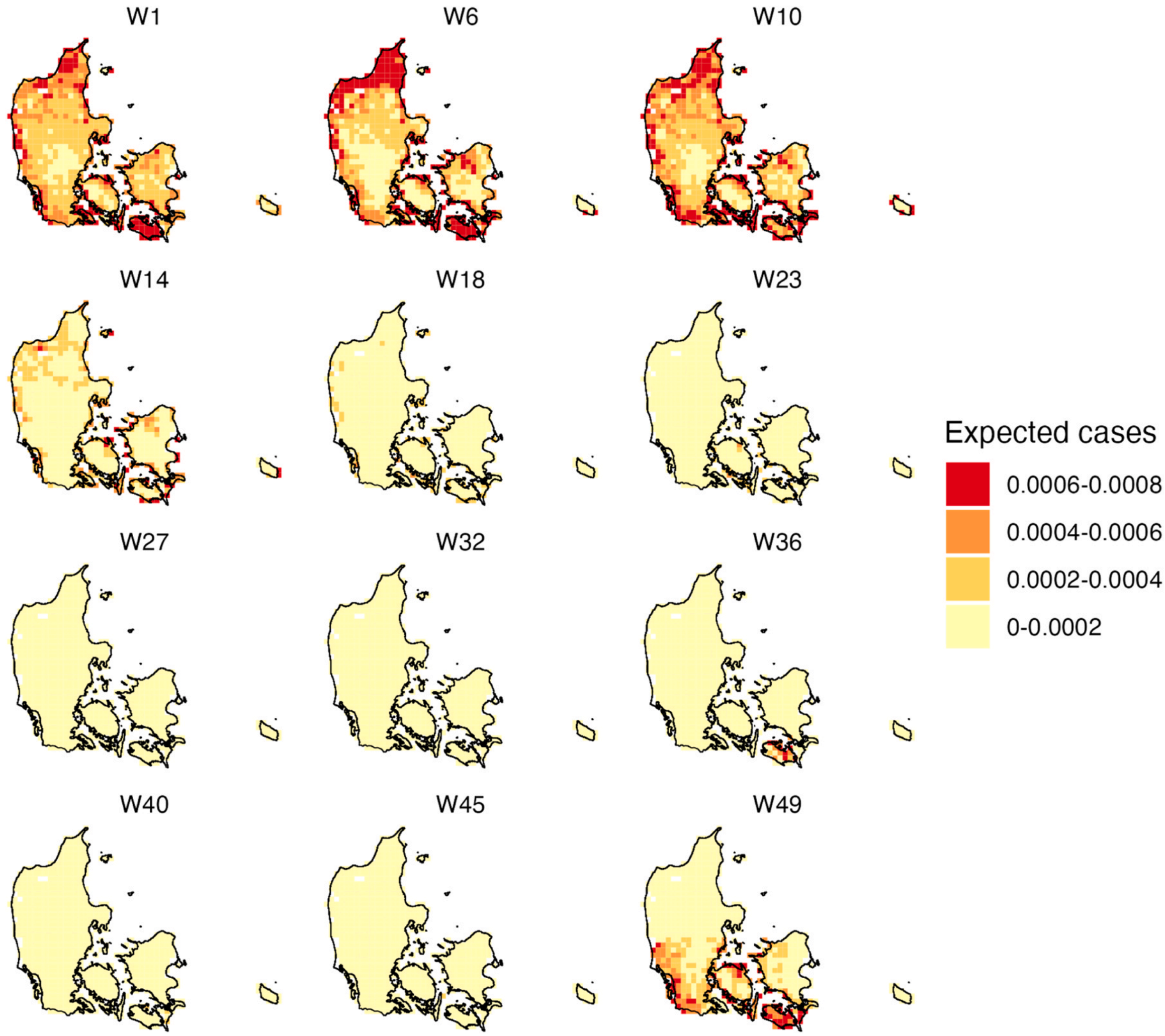

**Fig. 2.** Spatial distribution of expected infected cattle by wild birds across Denmark. The maps show the first week of each month in the year 2024.

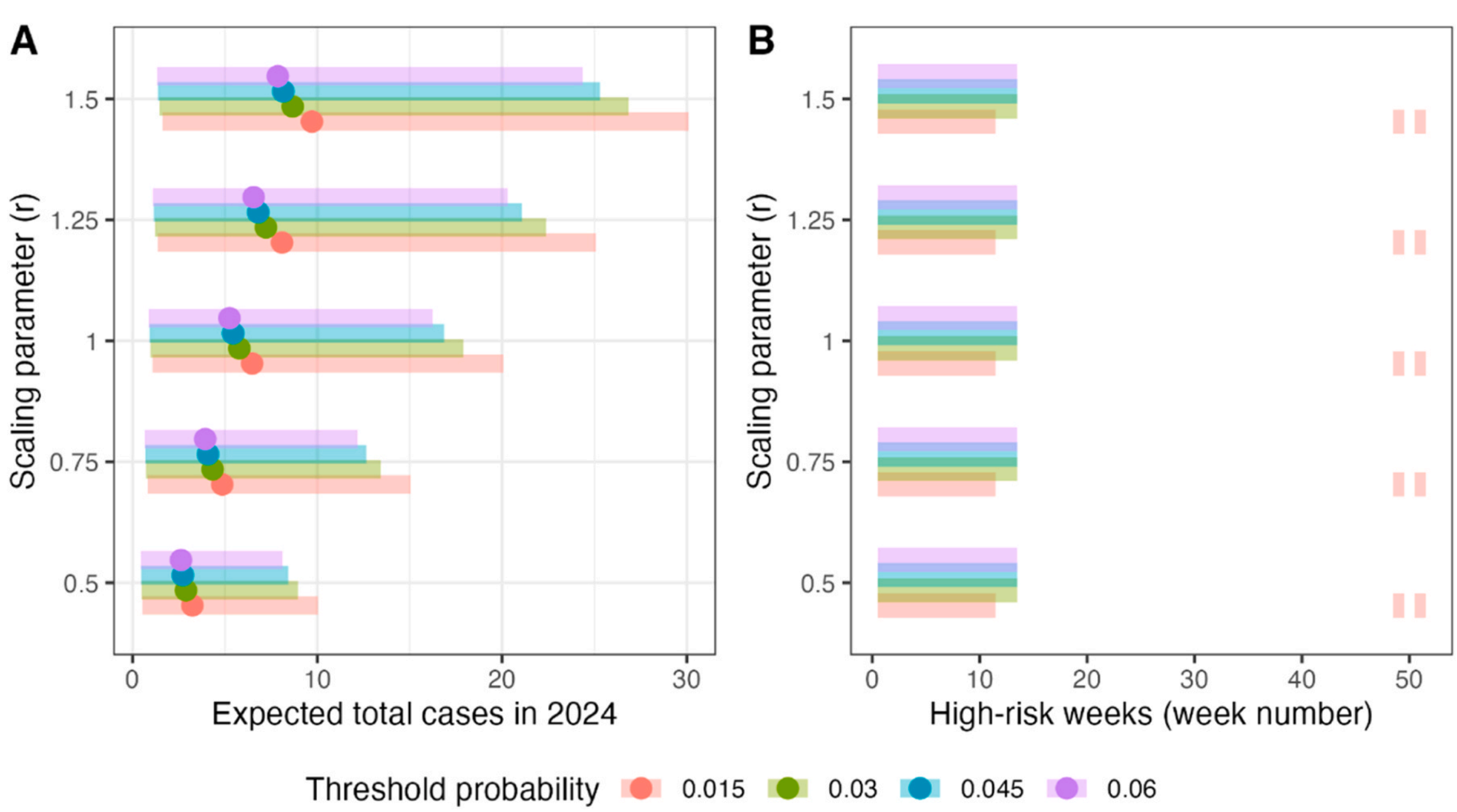

**Fig. 3.** Sensitivity analysis showing the impact of model parameters on (A) the total number of expected cases in 2024 (dots represent mean values; error bars represent 95% confidence intervals and (B) the predicted high-risk weeks (weeks falling within the top 25% risk quantile).

Both parameters, threshold probability and scaling parameter *r*, had negligible effects on the spatial distribution of predicted risk. The locations of high-risk areas (cells within the top 25% risk quantile) remained consistent across parameter settings (Supplementary Figure S4).

### 3.3. Structural sensitivity analysis

Results presented in Sections 3.1 and 3.2 are based on the frequency-dependent model structure. To assess the robustness of predictions to alternative transmission assumptions, we evaluated a density-dependent transmission model, re-estimated the transmission rate parameter, and extrapolated predictions for Denmark.

Under the density-dependent model, the weekly transmission rate from wild birds to cattle ($\beta_{DK_den} \cdot Prev_{DK}$) was estimated as $6.17 \times 10^{-8}$ (95% CI: $1.02 \times 10^{-8}$, $1.9 \times 10^{-7}$).

Thus, the expected number of spillover infections per cell per week in Denmark, ranging from 0 to $3.3 \times 10^{-2}$, with a median of $7.6 \times 10^{-4}$. This corresponds to cumulative expected number of spillover cases in Denmark of 19.7 (95% CI: 3.27, 60.8) in 2024, and 6.26 (95% CI: 1.04, 19.3) during the first eight months of 2025. Consequently, the annual number of expected cases was 3.6 times higher than that predicted by the frequency-dependent model.

The temporal pattern of predicted risk was similar to that obtained under the frequency-dependent model assumption, exhibiting peaks at the start and end of the year that align with wild bird migration periods (Supplementary Figure S5). In contrast, the spatial patterns differed between the two model structures (Fig. 4). Under the frequency-dependent assumption (Fig. 4A), spillover risk was primarily driven by the relative abundance of wild birds per cattle, resulting in higher predicted risk along coastal regions, particularly in northern and eastern Denmark. By comparison, the density-dependent model (Fig. 4B), predicted elevated risk in areas with high cattle densities, particularly along west and south coastline of Jutland and near the German border. The difference map (Fig. 4C) highlights these spatial contrasts; coastal areas identified as high risk under the frequency-dependent model often shift to lower quartiles under the density-dependent model, whereas inland regions with dense dairy cattle populations increase in risk classification.

## 4. Discussion

While current surveillance in Denmark primarily targets poultry and wild birds, active surveillance or retrospective serological studies in cattle have not yet been implemented. Consequently, the extent of possible HPAIV exposure in cattle remains unknown. To support outbreak preparedness, we developed a quantitative risk assessment tool to estimate the spillover risk of HPAIV from wild birds to cattle in Denmark. The tool combines information on HPAIV presence in wild birds and wild bird abundance from existing platforms such as Bird Flu Radar and eBird, and was calibrated using observed spillover events data in U.S. cattle before being applied to the Danish cattle population. This study provides a quantitative assessments of HPAIV spillover to cattle, addressing a gap noted in previous reviews (Pinheiro Marques et al., 2025).

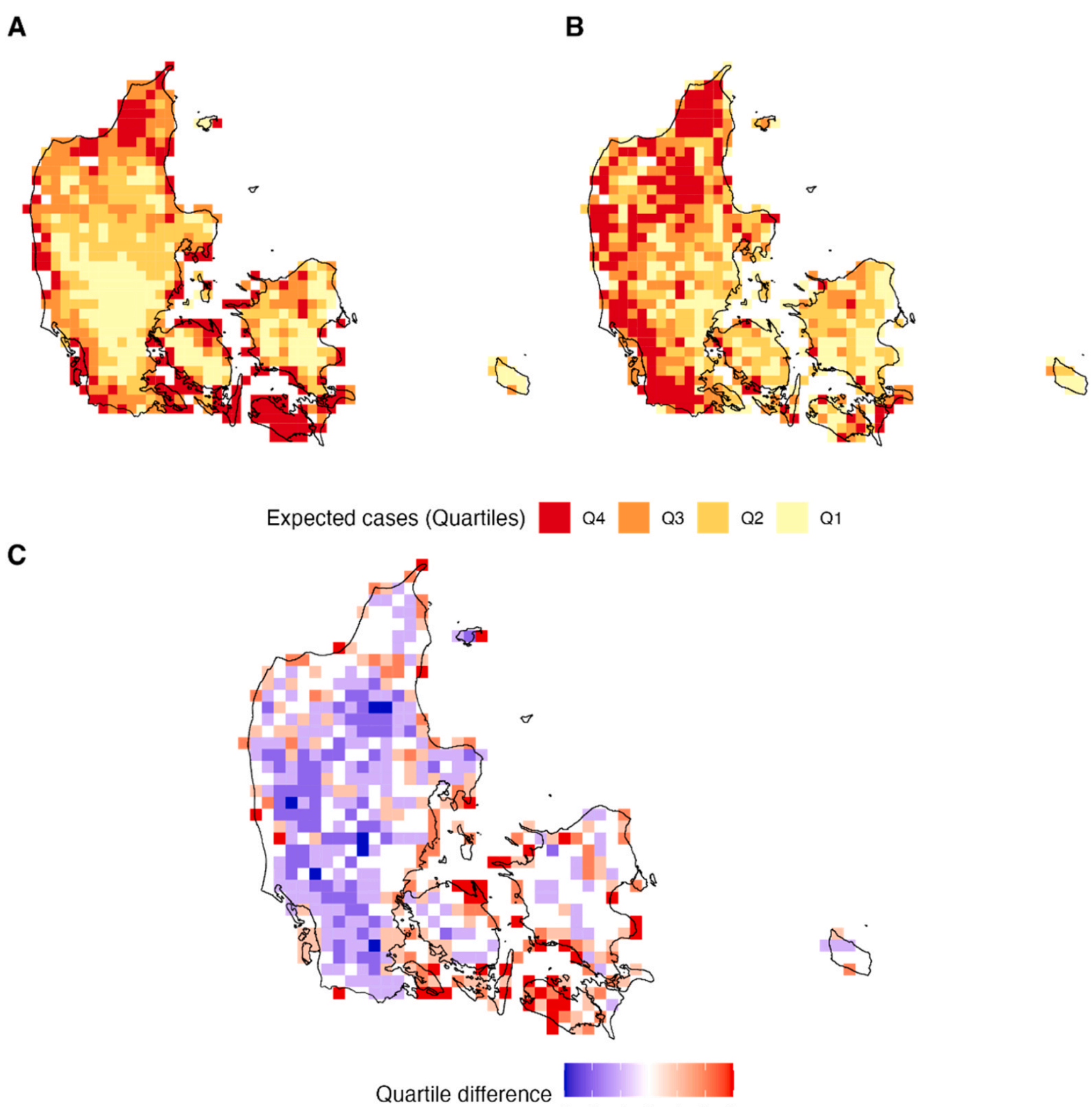

**Fig. 4.** The spillover risk maps for Denmark: A*)* Risk map under the frequency–dependent model assumption. Each cell is classified into quartiles (Q1–Q4) based on the expected number of cases in 2024 in Denmark; B) Risk map under the density-dependent model assumption; C) Heatmap of differences in quartile classification between the two model assumptions. Values indicate the quartile under the frequency–dependent model minus the quartile under the density-dependent model, ranging from –3 to + 3.

The model, under both transmission formulation assumptions, identified broadly similar high-risk areas for spillover along the Danish coastline and near the German border. The relative emphasis differed between the two assumptions: frequency-dependence placed greater weight on areas with high wild-bird abundance, whereas density-dependence shifted higher risk slightly inland into regions with dense dairy farming. These differences indicate that spatial prioritization is sensitive to the assumed contact pattern between species, yet the areas consistently highlighted by both assumptions (e.g. northern Jutland) may represent robust candidates for targeted cattle surveillance. Given that observed HPAIV incursions in Danish poultry have occurred mainly near wetlands or in association with wild bird presence rather than in high-density production zones (Liang et al., 2025), a frequency-dependent assumption maybe be considered more plausible for wildlife-to-livestock spillover in the Danish context. More empirical data on wild bird-cattle interactions will be important for understanding which contact structure better reflects spillover processes.

Temporally, the highest risk was predicted between December and March, coinciding with the seasonal migration of wild birds. Throughout the year, weekly expected spillover events remained below 0.35 cases, indicating that the likelihood of introduction in any given week is small. This pattern aligns with the observations in European countries where spillovers to livestock have been rarely observed (WOAH, 2025) with only isolated events (Banyard et al., 2025; Wageningen Bioveterinary Research, 2026). Although weekly hazards are small, they recur over time, and the sum of these weekly expectations yields an annual expectation of 5.4 introductions in 2024. This suggests that spillover events cannot be ruled out over longer time horizons, even if the weekly probabilities are low. Whether any introduced infections in Danish cattle would simply fade out or could sustain within-herd and spread between-herd, as occurred in the U.S., remains unknown and needs further investigation.

Due to limited information on spillover from wild birds to cattle, the model relies on several assumptions that may influence the results and should be carefully considered when interpreting the presented findings. First, spillover events reported in the U.S. may be underreported, potentially leading to an underestimation of the transmission rate parameter. Second, we assumed $r = 1$ under the baseline scenario, representing comparable infection pressure per between-species contact in the U.S. and Denmark. This assumption reflects limited information on the relative HPAIV prevalence in wild birds during outbreaks ($\frac{Prev_{DK}}{Prev_{US}}$) and on potential differences in transmission rate parameters ($\frac{\beta_{DK}}{\beta_{US}}$). To better characterize how uncertainty in $r$ influences model predictions, we first varied $r$ in sensitivity analysis and then performed a regression analysis on the predicted cases. The regression showed a linear relationship between $r$ and the predicted cases, meaning that uncertainty in $r$ primarily affects the absolute magnitude of predicted cases, while temporal and spatial patterns remain consistent. This property ensures that the model can be easily rescaled once more data become available. Part of this uncertainty in ($\frac{\beta_{DK}}{\beta_{US}}$) likely reflects differences in cattle husbandry and biosecurity practices between two countries. Unlike Nevada and Arizona in the U.S., where the spillovers took place, Danish dairy cows are predominantly housed indoors all year round with less than 30% on pasture during summer time (Annon, 2021). This may reduce contact with wild birds and contaminated outdoor environment and thereby lower the spillover risk. However, large numbers of starlings have been observed entering stables in spring and autumn (Budde et al., 2025; Sams, 2025), showing higher-than-expected contact between species in indoor systems. Although starlings exhibit limited intraspecies transmission and are unlikely to act as efficient biological hosts, they may transiently carry or transport contaminated material (Ellis et al., 2021). Future research should investigate the prevalence of HPAIV infection in these wild birds and examine how different cattle management systems (indoor vs outdoor) influence the contact between wild birds and cattle.

As our model relies on Bird Flu Radar to indicate the presence of HPAIV in wild birds, it inherits several limitations from Bird Flu Radar (Gargallo et al., 2022). First, Bird Flu Radar only accounts for bird movements and outbreak data within the EU and the UK. As a result, infections occurring in migratory birds originating from regions outside Europe such as Russia or Siberia are only captured once cases are detected within the EU, which may introduce a delay in capturing presence of HPAIV in wild birds. Second, Bird Flu Radar focuses on limited waterfowl species (order *Anseriformes*) to predict the HPAIV presence in wild birds, as they are conventionally understood as the main reservoirs of HPAIV in Europe and increases in incidence of HPAIV in waterfowl correlate with increased incidence in outbreaks of HPAIV in poultry particularly in the autumn and winter months. However, in recent years there are H5Nx HPAIV genotypes adapted to and circulating among seabirds (i.e., gulls and terns from order *Charadriiformes*), with higher incidence during summer months (EFSA et al., 2025; Indykiewicz et al., 2025; Lean et al., 2024; Rijks et al., 2022). While seabird abundance data can be obtained from eBird, their migratory patterns are not currently captured by the Bird Flu Radar and were therefore not incorporated into the spillover model presented in this study. Incorporating gulls and terns could increase estimated risks during the spring and summer months. Lastly, Bird Flu Radar includes fewer bird species than those used in our estimate of wild bird abundance, which may introduce some bias. To account for this potential mismatch, we tested alternative threshold probabilities in the Bird Flu Radar tool. Lowering the threshold probability can increase the sensitivity for detecting HPAIV presence and can therefore partially mitigate the bias arising from the limited number of species considered in Bird Flu Radar.

An important question is whether the predicted spillover events reflect an over- or under-estimation of the true spillover risk. The direction of bias is difficult to determine given current knowledge gaps and data limitations. Several factors could lead to underestimation, including underreporting of spillover events in U.S., limited inclusion of wild bird species and the exclusion of transboundary introduction pathways. In particular, the model focuses on local infectious pressure from wild birds present in Denmark and does not consider potential virus introduction from the United States via importation of contaminated animal products, even though this risk has been assessed as highly unlikely (EFSA, Alvarez, et al., 2025). In contrast, the absence of reported spillover events in Danish cattle to date, could suggest an overestimation of spillover risk by the model; however, this may also reflect limited surveillance rather than a true absence of spillover. Furthermore, the model assumes cattle remain fully susceptible throughout the study period and does not explicitly account for loss of susceptible due to within-herd transmission, which could reduce the number of animals available for additional spillover events if sustained cow-to-cow transmission were to occur. Uncertainties related to cattle husbandry practices and biosecurity, could contribute to either over- or under-estimation of spillover risk. Taken together, these considerations indicate that the absolute predicted spillover events is subject to uncertainty in both directions, and that addressing the knowledge gaps identified above will be essential for improving understanding of HPAIV spillover risk in cattle.

This study focuses on the risk of spillover into the Danish dairy cattle population, as the first stage, while potential onward transmission between Danish farms represents the second stage to assess requiring consideration of livestock movement patterns, and farm management practices, including biosecurity standards in Denmark (Chang et al., 2025; Conrady et al., 2023, 2024; Rawson et al., 2025). Whether HPAIV could establish sustained transmission within the Danish cattle industry ultimately depends on whether the between-herd transmission potential (between herd $R_0$ exceeds 1 or not). In U.S., the movement of lactating cows between farms has likely facilitated effective between-herd transmission. The extent to which similar movements could contribute to onward spread in Denmark remains unclear and needs to be investigated. Future studies could use the predicted introduction cases from this work to assess whether such events might establish or maintain between cattle herd transmission in Denmark.

This study provides a quantitative risk assessment for the Danish cattle sector, offering a framework to guide targeted surveillance and to optimize resource allocation in high-risk areas and periods identified in this study. Validation of the model through targeted surveillance in these high-risk areas and periods (e.g., through milk-based testing) would further strengthen confidence in the predictions and help refine preparedness strategies. While tailored to Denmark, the presented modeling framework is flexible and can be adapted for use in other countries, supporting broader regional or European applications.

## 5. Conclusion

This study predicted robust spatial and temporal patterns, showing elevated risk during winter and early spring and in coastal regions and areas near the German border. While uncertainty in the scaling parameter affects the absolute number of expected cases, it does not alter these patterns, suggesting that the model provides a reliable basis for prioritizing surveillance efforts. By focusing targeted surveillance on high-risk regions and seasons, these findings can help authorities and industry stakeholders allocate resources efficiently and improve early detection of potential cross-species transmission.

## CRediT authorship contribution statement

**You Chang:** Writing – review & editing, Writing – original draft, Visualization, Validation, Software, Methodology, Investigation, Funding acquisition, Formal analysis, Data curation, Conceptualization. **Beate Conrady:** Writing – review & editing, Writing – original draft, Validation, Resources, Project administration, Methodology, Investigation, Funding acquisition. **Jose L. Gonzales:** Writing – review & editing, Writing – original draft, Validation, Methodology. **Erik Rattenborg:** Writing – review & editing, Writing – original draft, Resources, Methodology, Data curation. **Mart C.M. de Jong:** Writing – review & editing, Writing – original draft, Validation, Methodology.

## Declaration of Competing Interest

The authors declare that they have no known conflict of financial interests or personal relationships that could have appeared to influence the work reported in this paper.

## Acknowledgement

This project is funded by Danish Milk Levy fund (Mælkeafgiftsfonden)and Cattle Levy fund (Kvægafgiftsfonden).

## Appendix A. Supporting information

Supplementary data associated with this article can be found in the online version at doi:10.1016/j.prevetmed.2026.106844.